\documentclass{iopart}
\usepackage{iopams}
\usepackage{graphicx}
\begin{document}

\title{Two simple systems with cold atoms: quantum chaos tests and nonequilibrium dynamics}

\author{Cavan Stone$^1$, Yassine Ait El Aoud$^1$, Vladimir A Yurovsky$^2$ and Maxim Olshanii$^1$}
\address{$^1$Department of Physics, University of Massachusetts Boston, Boston MA 02125, USA}
\address{$^2$School of Chemistry, Tel Aviv University, 69978 Tel Aviv, Israel}
\ead{Maxim.Olshanii@umb.edu}


\date{\today}

\begin{abstract}
This article is an attempt to provide a link
between the quantum nonequilibrium dynamics of cold gases
and fifty years of progress in
the low-dimensional quantum chaos. We identify two atomic systems lying
on the interface: two interacting atoms in a harmonic multimode waveguide and an
interacting two-component Bose-Bose mixture in a double-well potential.
In particular, we study the level spacing distribution, the wavefunction statistics,
the eigenstate thermalization, and the ability to thermalize in a relaxation process as such.
\end{abstract}

\maketitle




\section{Introduction}
In this article, we identify two simple atomic system lying on the interface between the nonequilibrium
dynamics of cold gases and low-dimensional
quantum chaos. The system of two short-range-interacting atoms in a harmonic waveguide is compared to
the well studied \v{S}eba billiard \cite{seba1990}. A two-component interacting Bose-Bose mixture in a double-well
potential provides a realization of a quantum particle moving on a four-dimensional surface of a tensor product
of two spheres under a non-integrable perturbation. 

The waveguide problem allows for an exact analytic solution, in spite of the absence of a complete set of integrals
of motion. The double-well problem reduces to a numerical diagonalization of a small, a-few-hundred-by-a-few-hundred
matrix, a few minute calculation on a moderate laptop.

To assess the degree of quantum chaos, we utilize two standard quantum-chaotic measures: level spacing statistics \cite{bohigas1991,guhr1998}
and the statistics of the values of the wavefunction in the coordinate representation \cite{berry1977,heller1984}.
The former predicts that the spectra of quantum-chaotic systems have the statistical properties
of the spectra of random matrices. The latter dictates the Gaussian statistics for the wavefunction.

From the point of view of the nonequilibrium dynamics, we study both the degree of the eigenstate thermalization 
and the actual thermalization in an expansion from a class of realistic initial 
states. The eigenstate thermalization \cite{shnirelman1974,feingold1986,
flambaum1997,deutsch1991,srednicki1994}
-- the suppression of the 
eigenstate-by-eigenstate variance of quantum expectation values 
of simple observables -- provides an ultimate upper bound for 
a possible deviation of the relaxed value of an observable 
from its thermodynamical expectation, for any initial state in principle. 
However, recently a new direction of research has emerged:
quantum quench 
in many-body interacting systems \cite{berman2004,calabrese2007,flambaum2001,kollath2007,
manmana2007,rigol2008,sengupta2004,polkovnikov2008,kinoshita2006,hofferberth2007,hofferberth2008}.
In this class of problems, the initial state is inevitably decomposed into a large superposition
of the eigenstates of the Hamiltonian governing the dynamics, and the discrepancy between 
the result of the relaxation and thermal values is expected to be diminished.

Note that a related comprehensive study of the relationship between 
the quantum chaos and thermalization for atoms in optical lattices has been performed in Ref.\ \cite{santos2009}.




\section{Two bosons in a circular, transversely harmonic multimode waveguide}

In the case of the relative motion of two short-range-interacting
atoms in a circular, transversely harmonic waveguide \cite{yurovsky2010} , 
the full Hamiltonian
\begin{eqnarray}
\hat{H} = \hat{H}_{WG} + \hat{V}_{FH}
\label{H_2b}
\end{eqnarray}
can be split onto an integrable, ``free motion'' part and a non-integrable perturbation.
The unperturbed Hamiltonian is given by the sum of the longitudinal and  transverse
kinetic energies and the  transverse trapping energy $\hat{U} = \mu\omega_{\perp}^2\rho^2/2$,
\begin{eqnarray}
\hat{H}_{WG} = -{\hbar^2 \over 2\mu}{\partial^2\over\partial z^2}
-{\hbar^2\over 2\mu} \Delta_{\rho}+ \hat{U} - \hbar\omega_{\perp}
\quad,
\label{H0_2b}
\end{eqnarray}
where
$\omega_{\perp}$ is the transverse frequency,
$\Delta_{\rho}$ is the transverse two-dimensional Laplacian,
$\mu = m/2$ is the reduced mass, and $m$ is the atomic mass.
We will assume periodic boundary conditions along $z$, with a period $L$. In what follows, we will restrict the Hilbert space to 
the states of zero $z$-component of the angular momentum and even under the $z \leftrightarrow -z$ reflection. Note
that the zero-range interaction has no effect on the rest of the Hilbert space. 
The non-interacting eigenstates are products of the transverse two-dimensional zero-angular-momentum harmonic wavefunctions, labeled by the quantum number $n \geq 0$,
and the symmetric plane waves $\cos(2\pi l z/L)$, $l\geq 0$.
The unperturbed spectrum is therefore given by
$
E_{nl} = 2 \hbar \omega_{\perp} n + \hbar^2 (2\pi l/L)^2/(2\mu)
$

The interaction $\hat{V}_{FH}$ in (\ref{H0_2b}) couples the transverse and longitudinal degrees
of freedom. We assume the interaction to be of the Fermi-Huang type. 
It can be symbolically represented as a separable rank I perturbation \cite{albeverio2000}:
\begin{eqnarray}
\hat{V}_{FH} = V|{\cal L}\rangle\langle {\cal R}|,
\label{V_2b}
\end{eqnarray}
with the formfactors $|{\cal L}\rangle=\delta_{3}({\bi r})$ and $\langle {\cal R}|=\delta_{3}({\bi r}) (\partial/\partial r)(r\,\cdot )$.
Here, 
the interaction strength is $V=(2\pi\hbar^2 a_{s}/\mu)$, where
$a_{s}$ is the three-dimensional $s$-wave scattering length.
The rigorous definition of the operator (\ref{V_2b}) reads
\begin{eqnarray}
\langle \chi | \hat{V}_{FH} | \psi \rangle \equiv 
V \chi^{\star}({\bi r} = {\bf 0}) \lim_{r \to 0} \frac{\partial}{\partial r} (r \psi(r,\,\theta,\,\phi))
\quad.
\label{V_2b_rigorous}
\end{eqnarray}
Following the the Refs.\ \cite{albeverio2000,seba1990}, it can be shown that the operator 
(\ref{V_2b_rigorous}) produces a member of a (parametrized by $V$) family of self-adjoint extensions 
of the ``free motion'' Hamiltonian (\ref{H0_2b}). 

Note that the model (\ref{H_2b}), with an unbounded along $z$ motion, was first used to derive  
the effective one-dimensional atom-atom interaction in a monomode atom guide \cite{olshanii1998}. It was further 
extended to the multimode regime in Ref.\ \cite{moore2004}, where the intermode kinetic coefficients 
were derived. In this article, we consider a bound-motion analogue of the multimode guide of Ref.\ \cite{moore2004},
suitable for a study of the relaxation dynamics and chaotic properties.

Even though the perturbation 
of the form (\ref{V_2b}) destroys the complete set of integrals of motion of the unperturbed
system, the problem can be solved exactly. The solution for the full system can be obtained as follows. 
The action of the 
perturbation on each wavefunction produces the left formfactor $|{\cal L}\rangle$.
As a result, any eigenstate $|\alpha\rangle$ of the perturbed system 
functionally coincides with the Green function of the 
unperturbed Hamiltonian taken at the energy of the eigenstate $E_{\alpha}$. 
The following eigenenergy equation then applies, along with 
an expression for the eigenstates:
\begin{equation}
\sum\limits_{\vec{n}}
{\langle {\cal R}|\vec{n}\rangle \langle \vec{n} |{\cal L} \rangle
 \over E_{\alpha}-E_{\vec{n}}} =\frac{1}{V},
\quad
|\alpha\rangle \propto \sum\limits_{\vec{n}}
{|\vec{n}\rangle \langle \vec{n} |{\cal L} \rangle
 \over E_{\alpha}-E_{\vec{n}}},
\label{psiGreen}
\end{equation}
where $|\vec{n}\rangle$ stands for an  eigenstate of the unperturbed system of energy $E_{\vec{n}}$.
Similar expressions were obtained in Refs.\
\cite{seba1990,albeverio1991} for the case of the \v{S}eba billiard and its generalizations.
Solutions of this type, for the integrable
case of a spherically symmetric harmonic confinement, are presented 
in Ref.\ \cite{busch1998}.

Substituting the above non-interacting eigenstates and eigenenergies to the Eq.\ (\ref{psiGreen}) and summing over $l$, one obtains 
the following expression for the interacting eigenstates:
\begin{eqnarray}
\langle \rho ,z | \alpha\rangle =
C_{\alpha} \sum\limits^{\infty
 }_{n=0}{\cos\left( 2\sqrt{\epsilon _{\alpha}-\lambda n}\zeta \right)
 \over \sqrt{\epsilon _{\alpha}-\lambda n}\sin\sqrt{\epsilon _{\alpha}-\lambda
 n}} e^{-\xi/2} L^{(0)}_{n}\!\left(\xi\right) 
,
\label{psieps}
\end{eqnarray}
where the rescaled energy $\epsilon_{\alpha}$ is given by $\epsilon_{\alpha}\equiv\lambda E/\left(
 2\hbar\omega_{\perp }\right)$, $\lambda \equiv\left( L
/a_{\perp }\right) ^{2}$  is the aspect ratio,
$\zeta \equiv  z/L-1/2$, \,$\xi \equiv \left(\rho/a_{\perp }\right)^{2}$, \,   
$L^{(0)}_{n}(\xi)$ are the Legendre polynomials, and 
$a_{\perp }=\left(\hbar/\mu\omega _{\perp}\right)^{1/2}$ is the size of the 
transverse ground state. 
The normalization constant $C_{\alpha} $ is expressed as
\begin{eqnarray}
C_{\alpha} ={2\over a_{\perp }\sqrt{\pi L}}
\biggl\lbrack \sum\limits^{\infty }_{n=0}\biggl
({\cot\sqrt{\epsilon_{\alpha} -\lambda n}\over \left( \epsilon_{\alpha} -\lambda
 n\right) { } ^{3/2}}+ 
{1\over \left( \epsilon_{\alpha} -\lambda n\right) \sin^{2}\sqrt{\epsilon_{\alpha}
 -\lambda n}}\biggr)\biggr\rbrack ^{-1/2} . \label{Ceps}
\end{eqnarray}

Extracting the regular part of the wavefunction (\ref{psieps}) via 
the procedure  given in Refs.\ \cite{moore2004,yurovsky2008b}, we arrive at the
following transcendental equation for the eigenenergies $\epsilon_{\alpha}$:
\begin{equation}
\sqrt{\lambda}\sum\limits^{\infty }_{n=0}
{\cot\sqrt{\epsilon_{\alpha} -\lambda n}+i\over \sqrt{\epsilon_\alpha
 -\lambda n}}-\zeta \left( {1\over 2},-{\epsilon_{\alpha} \over \lambda
 }\right) ={a_{\perp }\over a_{s}} 
\quad, 
\label{Eqeps}
\end{equation}
where $\zeta \left( \nu ,x\right)$ is the Hurwitz $\zeta $-function
(see \cite{yurovsky2008b}). The summands in the sums Eqs.\ (\ref{psieps}), (\ref{Ceps}), and (\ref{Eqeps}) 
decay exponentially with $n$, leading to the fast converging series.
Note that the imaginary parts of the two terms in the left hand side of the (\ref{Eqeps})
cancel each other automatically. Similar solutions were obtained for
two atoms with a zero-range interaction in a cylindrically-symmetric
harmonic potential \cite{Idziaszek2005}
(that system was analyzed numerically in Ref.\ \cite{bolda2003}).
Higher
partial wave scatterers
were analyzed in the Ref.\
\cite{Kanjilal2007}.

At rational values of $\lambda/\pi^2$ 
the unperturbed energy spectrum shows degeneracies that
are not fully lifted in the full [deduced from (\ref{Eqeps})] spectrum.
To minimize the effect of the degeneracies, we,
following Ref.\ \cite{seba1990},
fix the length of the guide by $(L/a_{\perp})^2 \equiv \lambda = 2 \phi_{gr} \pi^7 \approx 9774$,
where $\phi_{gr} = (1+\sqrt{5})/2$ is the golden ratio. 




\section{A two-component interacting Bose gas in two coupled potential wells}
Next, we are going to consider a two-component
Bose-Bose mixture trapped in a two-well potential. 

The second-quantized Hamiltonian is 
\begin{eqnarray}
&&
\hat{H} = \hat{H}_{0} + \hat{V}
\label{two_well_hamiltonian}
\\
&&
\hat{H}_{0} = \frac{1}{2} \sum_{i=a,\,b} \sum_{j=a,\,b} \left(
                       U_{ijL} \hat{c}^{\dagger}_{iL} \hat{c}^{\dagger}_{jL} \hat{c}^{}_{iL} \hat{c}^{}_{jL} 
                     + U_{ijR} \hat{c}^{\dagger}_{iR} \hat{c}^{\dagger}_{jR} \hat{c}^{}_{iR} \hat{c}^{}_{jR}
							\right)
\label{two_well_H0}
\\
&&
\hat{V} = -2J \sum_{i=a,\,b} (\hat{c}^{\dagger}_{iL} \hat{c}^{}_{iR} + \hat{c}^{\dagger}_{iR} \hat{c}^{}_{iL})
\label{two_well_V}
\quad,
\end{eqnarray}
where $\hat{c}^{\dagger}_{iL}$($\hat{c}^{\dagger}_{iR}$) is the creation operator for the atom of the $i$-th type in the left(right) well,
$a$ and $b$ denote the two species,
and the factor of 
$-2$ in the definition of $V$ is used to ensure the consistency with the conventional 
definition of the hopping constant in the theory of multi-well lattices with periodic boundary conditions.

We fix the coupling constants $U_{ijL(R)}$ to 
$
U_{aaR} = 0.3041\, U,\,
U_{abR} = 1.0410\, U,\,
U_{bbR} = 0.3684\, U,\,
U_{aaL} = 0.3085\, U,\,
U_{abL} = 1.4574\, U,\,
U_{bbL} = 0.4524\, U
$, where $U$ is the overall coupling strength. The relative strengths of the interaction were chosen once
and for all
from a uniform random distribution between $0$ and $2$. The same set was used in all further numerical experiments.

The set of deviations of the number of atoms of a given type in the right well from the half of the total number of atoms of 
this type constitutes a convenient set of quantum numbers: $s_{a} \equiv N_{aR} - N_{a}/2,\, s_{b} \equiv N_{bR} - N_{b}/2$,
where $\hat{N}_{aR} \equiv  \hat{c}^{\dagger}_{aR} \hat{c}^{}_{aR}$ and $\hat{N}_{bR} \equiv  \hat{c}^{\dagger}_{bR} \hat{c}^{}_{bR}$. The 
observable $\hat{s}_{a}$ has been chosen to be the principal observable of interest in the eigenstate thermalization
and relaxation studies. Below, we fix the numbers of atoms to $N_{a}=N_{b}=N=20$.

Analogously to the one-specie case \cite{milburn1997}, the Hamiltonian (\ref{two_well_hamiltonian}) is integrable in both $J \gg U$ and $J \ll U$ limits.

In the $J \ll U$ limit, the eigenstates are represented by the Fock states (eigenstates of $\hat{H}_{0}$, see (\ref{two_well_H0})) 
where number of bosons of each type 
in each well is fixed. For our choice of the coupling constants, the energy as a function of $(s_{a},\,s_{b})$ constitutes a hyperboloid.
At low energies, the spectrum of $\hat{H}_{0}$ 
is dominated by the eigenstates where the atoms of different type are localized 
in the opposite wells: $(s_{a},\,s_{b}) \approx (\pm N,\,\mp N)$. At high energies, the atoms prefer to localize in the 
same well: $(s_{a},\,s_{b}) \approx (\pm N,\,\pm N)$.

In the opposite limit $J \gg U$, the good quantum numbers are given by the numbers 
of particles of each type in the even and odd  (with respect to the wells) one-body states.
In this limit, the mean-field Hamiltonian 
\begin{eqnarray}
\hat{H}_{\mbox{\scriptsize mf}} &\equiv& \hat{V} + \langle \hat{H}_{0} \rangle
\label{two_well_Hmf}
\\
&=& 
\hat{V} + 
\frac{N}{8} 
\left( (N-2)( U_{aaL}+U_{aaR}+U_{bbL}+U_{bbR} ) \right.
\nonumber
\\
&&
\qquad\qquad
\left.
+ 2N(U_{abL}+U_{abR})   \right)
\nonumber
\quad,
\end{eqnarray}
with the hopping as the principal component, generates a good approximation to the energy 
spectrum. The spectra of the full Hamiltonian $\hat{H}$ (\ref{two_well_hamiltonian}), the 
pure-interaction Hamiltonian $\hat{H}_{0}$ (\ref{two_well_H0}), and the pure-hopping-mean-field
Hamiltonian $\hat{H}_{\mbox{\scriptsize mf}}$ are shown at Fig.\ \ref{Fig_cavan_spectra_and_eth}a.
Note the degeneracies in the spectrum of the pure-hopping Hamiltonian $\hat{H}_{\mbox{\scriptsize mf}}$.
Recall that in our model, the hopping constants are the same for both types of atoms;
as a result, each energy level corresponding to the total of $N_{+}$ atoms in the even state 
$(1/\sqrt{2}) (|L\rangle + |R\rangle)$ is ($N-|N_{+}-N|+1$)-fold degenerate. 

The point $\alpha \approx 190$ where the exact, pure-interaction, and pure-hopping eigenenergy curves
coincide corresponds to the maximal depolarization, $\langle \hat{s}_{a(b)}\rangle \approx \langle \hat{c}^{\dagger}_{a(b)L} \hat{c}^{}_{a(b)R} \rangle \approx 0$,
where neither of the two integrable limiting models apply. 
This is the point where the motion is expected to be ``maximally chaotic''.

The attractive feature of the two-well system is that it can be mapped to a system consisting of 
a single particle moving on the four-dimensional surface
of a tensor product of the two two-dimensional spheres. The map is 
$| s_{a},\,s_{b} \rangle \to | (L_{a}\!=\!N_{a}/2,\,m_{a}\!=\!s_{a}),\,(L_{b}\!=\!N_{b}/2,\,m_{b}\!=\!s_{b}) \rangle $,
where $L_{a}$($L_{b}$) is the total angular momentum on the first(second) sphere, and $m_{a}$($m_{b}$) is the corresponding 
projection to the $z$-axis.
The corresponding Hamiltonian reads:
\begin{eqnarray}
&&
\hat{H} = 
\frac{1}{2}  (U_{aaL} \hat{{\cal L}}_{z,a}(\hat{{\cal L}}_{z,a}-1) + 2 U_{abL} \hat{{\cal L}}_{z,a}\hat{{\cal L}}_{z,b} + \ldots)
\label{two_sphere_hamiltonian}
\\
&&
\qquad
-
4J \sum_{i=a,\,b} \hat{{\cal L}}_{x,i}
\nonumber
\quad.
\end{eqnarray}
The total angular momenta $(\hat{\vec{\cal L}}_{a})^2 \equiv L_{a}(L_{a}+1)$ and $(\hat{\vec{\cal L}}_{b})^2 \equiv L_{b}(L_{b}+1)$ commute
with the Hamiltonian (\ref{two_sphere_hamiltonian}). In the many-boson picture, these momenta are related
to the number of bosons of each type as $L_{a(b)} = N_{a(b)}/2$ (both assumed to be even). For the one-specie case, similar mapping was employed 
in Ref.\ \cite{milburn1997}.

From the single-particle point of view,
the Hamiltonian (\ref{two_sphere_hamiltonian}) describes an external perturbation of the free motion along the bi-sphere. 
For clarity, 
we do not include the constant
``free'' kinetic energy $\propto (\hat{\vec{\cal L}}_{a})^2 + (\hat{\vec{\cal L}}_{b})^2$ in the Hamiltonian (\ref{two_sphere_hamiltonian}). 
 
In the coordinate representation, a given state 
\begin{eqnarray*}
|\psi\rangle = \sum_{s_{a},\,s_{b}} \langle s_{a},\,s_{b} | \psi \rangle | s_{a},\,s_{b} \rangle
\end{eqnarray*}
maps to a two-particle wavefunction (defined at the points of a tensor product of 
two unit spheres) given by
\begin{eqnarray}
&&
\psi(\theta_{a},\,\phi_{a},\,\theta_{b},\,\phi_{b}) = 
\label{sphere}
\\
&&\quad\quad
\sum_{s_{a},\,s_{b}} \langle s_{a},\,s_{b} | \psi \rangle \,
Y_{l\!=\!N_{a}/2,\,m\!=\!s_{a}}(\theta_{a},\,\phi_{a}) Y_{l\!=\!N_{b}/2,\,m\!=\!s_{b}}(\theta_{b},\,\phi_{b})
\quad,
\nonumber
\end{eqnarray}
where $Y_{l,\,m}(\theta,\,\phi)$ are spherical harmonics.

\begin{figure} 
\includegraphics[width=3.5in]{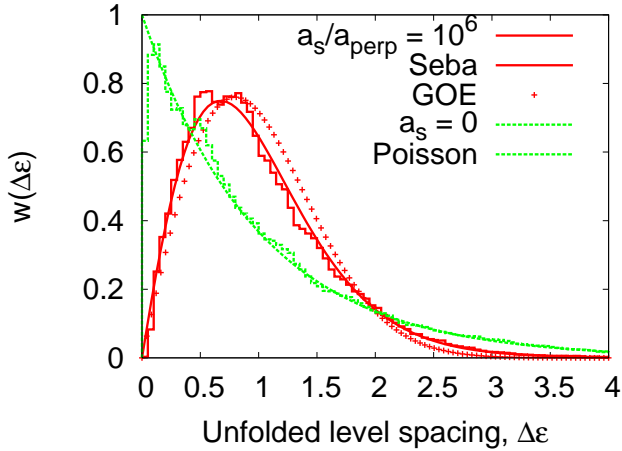}
\caption 
{
Two bosons in a multimode guide.
Nearest-neighbor spacing distribution averaged over $7\times 10^{6}$ eigenstates 
in the globally unitary regime $a_{s}/a_{\perp }=10^{6}$ and in the  
regime of no interactions. The Gaussian Orthogonal Ensemble prediction,
the \v{S}eba distribution, and the Poisson distribution are shown for comparison. 
Here and below, $\left(L/a_{\perp }\right)^{2} = 2 \phi_{gr} \pi^7 \approx 9774$.
}
\label{Fig_Volodya_level_statistics}
\end{figure}
\begin{figure}
\includegraphics[width=3.5in]{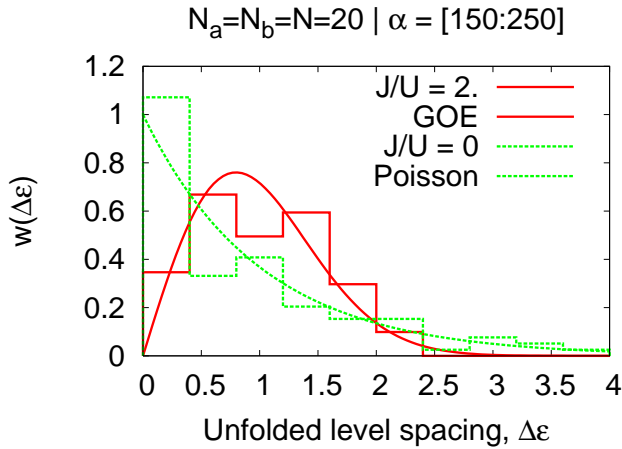}
\caption
{
Forty bosons in two wells.
Nearest-neighbor spacing distribution averaged over 100 eigenstates
in the ``maximally chaotic'' region $150 \le \alpha \le 250$.
Number of atoms of both type is $N_{a}=N_{b}=N=20$. 
The $J/U =2$ point 
corresponds to the empirically broadest fraction of the ``chaotic'' eigenstates.  
The case of no hopping is shown as well, along with the 
Gaussian Orthogonal Ensemble and Poisson predictions. Here and below,
the individual coupling constants are 
$
U_{aaR} = 0.3041\, U,\,
U_{abR} = 1.0410\, U,\,
U_{bbR} = 0.3684\, U,\,
U_{aaL} = 0.3085\, U,\,
U_{abL} = 1.4574\, U,\,
U_{bbL} = 0.4524\, U
$.
}
\label{Fig_Cavan_level_spacings_histogram}
\end{figure}
\begin{figure}
\includegraphics[width=4.5in]{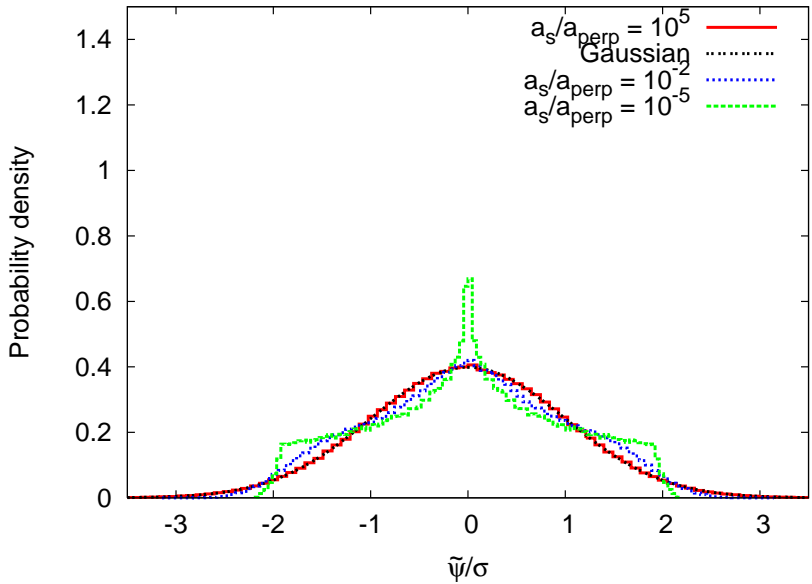}
\caption
{
Two bosons in a multimode guide.
The probability distribution for value of the eigenstate wavefunction
in the region close to the waveguide axis. The eigenstate 
number $67600$ corresponding to the energy
$E_{\alpha} \approx 4.24 \times 10^{6} \hbar\omega_{\perp}$ is shown.
Since the motion is restricted to the cylindrically symmetric 
states the reduced wavefunction $\tilde{\psi}(\rho,\,z) \equiv (1/\sqrt{2\pi\rho})\psi(\rho,\,z)$
was used.
The Gaussian distribution corresponds to the Berry conjecture.
The interaction strengths
$a_{s}/a_{\perp}=10^{5}$, $10^{-2}$, and 
$10^{-5}$ correspond to the unitary, intermediate, and non-interacting 
regimes respectively. 
The rest of the parameters is the same as  at Fig.\ \ref{Fig_Volodya_level_statistics}.
}
\label{Fig_Volodya_psi_berry_long}
\end{figure}
\begin{figure}
\includegraphics[width=4.5in]{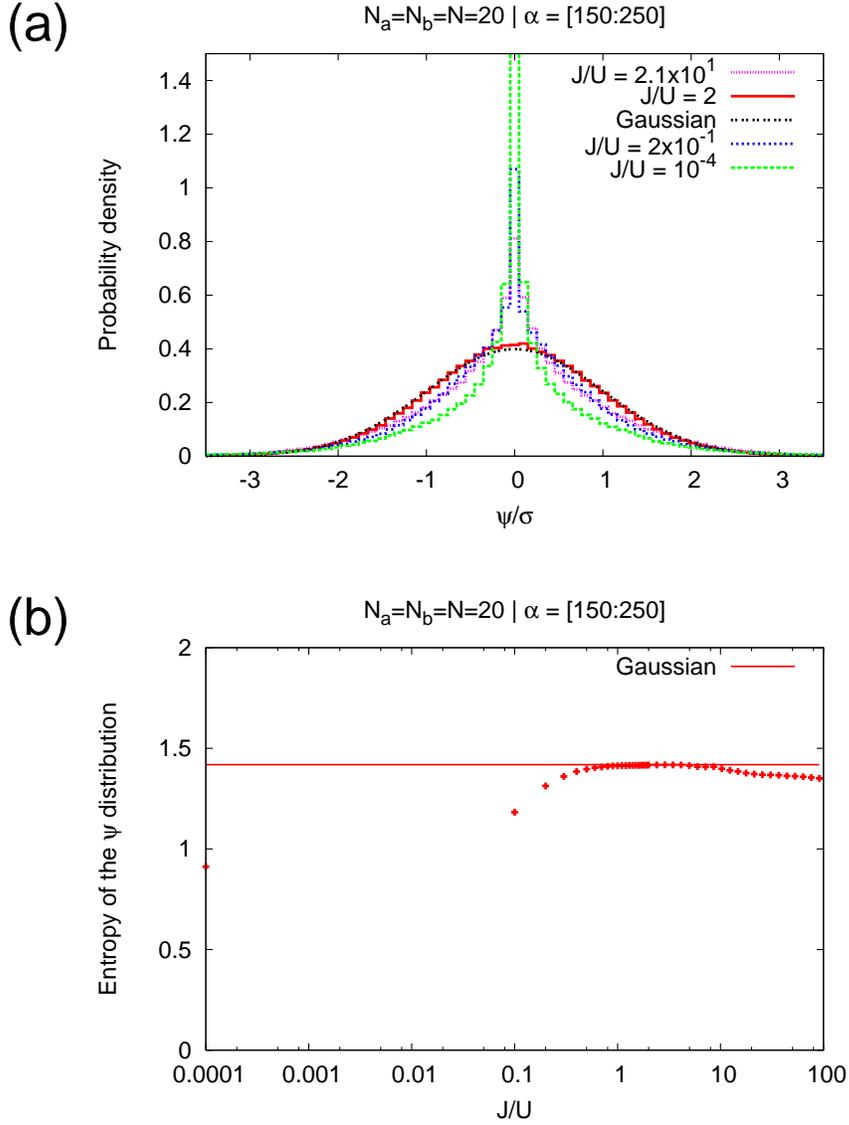}
\caption
{
Forty bosons in two wells.
(a)
The probability distribution for real part of the eigenstate wavefunction on
the whole surface of the bi-sphere. The distributions for the 
individual eigenstates were further averaged over an $150 \le \alpha \le 250$
window. The distribution is heavily peaked in both integrable limits, 
$J \ll U$ and $J \gg U$. 
(b) Entropy of the distribution as a function of the hopping constant.
The entropy converges to the Berry conjecture prediction in the region 
between $J/U \approx 1$ and $J/U \approx 5$. 
For both (a) and (b) the rest of the parameters is the same 
as for the Fig.\ \ref{Fig_Cavan_level_spacings_histogram}.
}
\label{Fig_Cavan_psi_gaussian}
\end{figure}
\begin{figure}
\begin{center}
\includegraphics[width=3.5in]{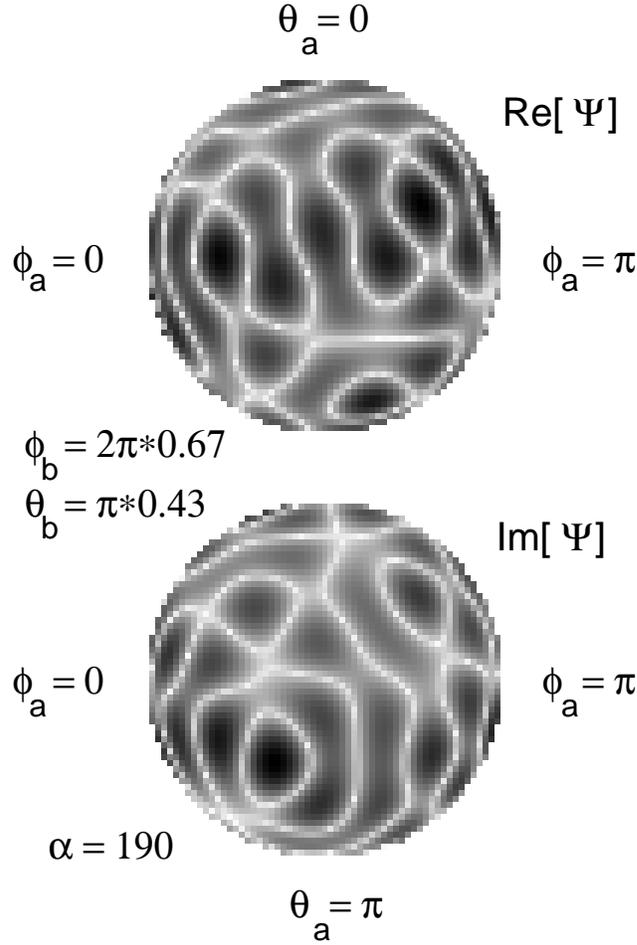}
\end{center}
\caption
{
Forty bosons in two wells.
The real and imaginary parts of the four-dimensional eigenstate wavefunction living on
the surface of a tensor product of two unit spheres. The ``maximally chaotic''
state $\alpha=190$ is used. The hopping is fixed to $J/U =2$. Even though 
in both $J \ll U$ and $J \gg U$ limits the eigenstates are strongly polarized,
for $J/U =2$ no preferential direction is visible.
The rest of the parameters is the same 
as for the Fig.\ \ref{Fig_Cavan_level_spacings_histogram}. To enhance contrast, the shading density
is made to be proportional to the forth power of the real and imaginary parts respectively.
}
\label{Fig_Cavan_psi_sphere}
\end{figure}
\begin{figure}
\begin{center}
\includegraphics[width=3.5in]{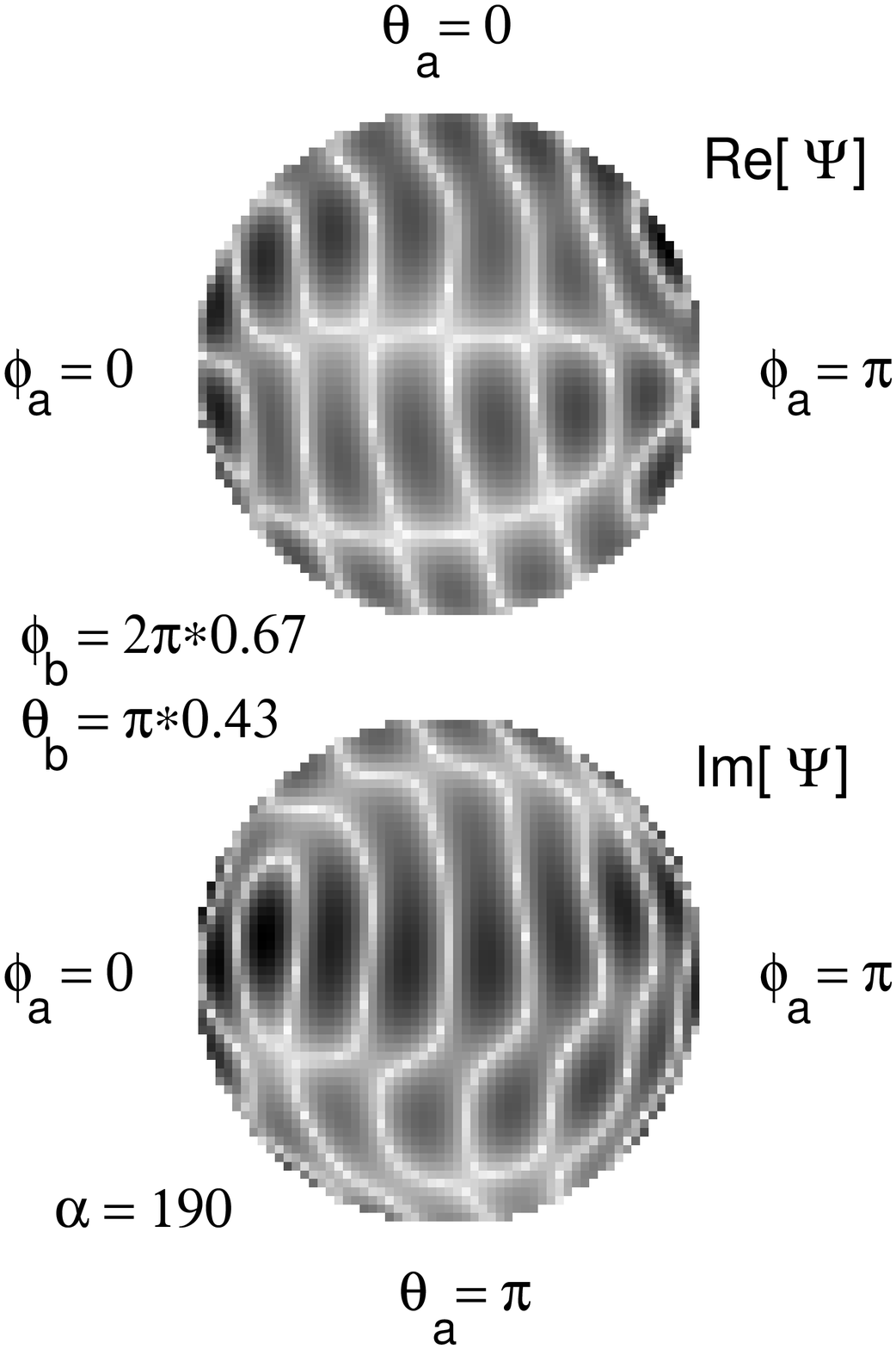} 
\end{center}
\caption
{
The same as on Fig.\ \ref{Fig_Cavan_psi_sphere} but in the regular regime, $J/U = .2$\,.
}
\label{Fig_Cavan_psi_sphere_noninteracting}
\end{figure}
%




\section{Standard quantum-chaotic tests}
%
\subsection{Level spacing statistics}
The commonly-used measure to access the degree of quantum chaos
the shape of level spacing distribution \cite{guhr1998}.
As usual, the starting point of the procedure is to 
determine the mean level density to be subsequently removed from the spectral data. The 
resulting energy spectrum---the so-called unfolded spectrum---does not contain any information specific 
to the system besides the dimensionless, homogeneous across the spectrum level spacing statistics.  
This statistics is the principal object of the investigation.

For the case of the waveguide, the mean level density was obtained from the semiclassical approximation to the unperturbed spectrum of the system.
Such a simplification was possible due to the fact that for a singular perturbation, 
the density of states is exactly the same for the interacting and non-interacting systems.
For the two-well system, the mean level density was approximated by a fit of the true energy spectrum to a 
seventh degree polynomial.

The integrable systems exhibit the Poisson statistics of the level spacings, with a peak at zero and an 
exponential decay at large spacings. One would expect to this type of statistics in a model the energy levels 
appear under no constraint, besides the given a priori mean level spacing. Indeed, in integrable systems,
the energy levels corresponding to different sets of (additional to the energy) integrals of motion 
can approach each other infinitely closely. It turns out that this property along is sufficient 
to ensure the Poisson statistics \cite{bohigas1991,guhr1998}.  

On the contrary, in a quantum system with no non-trivial integrals of motion, the energy levels tend to repel. 
This repulsion leads to the appearance of a hole, around zero spacing point, in the level spacing distribution. 
It is remarkable however that generally, quantum chaotic systems with real Hamiltonians exhibit exactly the 
same level spacing distribution, namely the one of the Wigner-Dyson type. An ensemble of orthogonal 
matrices with a Gaussian distribution of the matrix elements (the so-called Gaussian Orthogonal Ensemble (GOE)) 
serves as a paradigm for this class of Hamiltonians \cite{bohigas1991,guhr1998}. 

A multi-dimensional harmonic oscillator constitutes 
a remarkable exception from the integrable vs.\ chaotic dichotomy \cite{whan1997}.
The unfolded distribution
function in this case vanishes at the zero spacing and shows a sharp decay at large spacings.
Indeed, we have found that our waveguide system, with a harmonic confinement transversally and a periodic-boundary-conditions 
box longitudinally, is more sensitive to the effect of a non-integrable perturbation than a 
seemingly more elegant elongated three-dimensional harmonic oscillator; thus our preference for the former.  

The waveguide system shows systematic deviations from the 
GOE Random Matrix Theory predictions \cite{bohigas1991}.
Note that the unattainability of a complete quantum chaos
in singular billiards has been already addressed in the context 
of the \v{S}eba billiard -- flat two-dimensional rectangular billiard with a zero-range scatterer in the middle \cite{seba1990}.
Even though the \v{S}eba billiard shows some signatures of the
quantum-chaotic behavior, 
both the level statistics \cite{seba1991,albeverio1991,bogomolny2001}
and the momentum distributions in individual eigenstates \cite{berkolaiko2003} 
exhibit substantial deviations from the conventional quantum-chaotic behavior.
Our waveguide system belongs to the same class as the \v{S}eba billiard: it is represented by 
an integrable system perturbed by a separable rank-one
perturbation. For both systems, the eigenstates have the ``Green's function'' form (\ref{psiGreen}).  

The results of 
the study of the level spacing distribution 
in the waveguide system 
are fully consistent with the analogous results for the  
\v{S}eba billiard \cite{seba1991,albeverio1991}.
One can show that for $a_{s} \gg a_{\perp}$ (globally unitary regime), the scattering 
strength reaches the unitary limit for all eigenenergies of 
the full Hamiltonian.
In this regime the distribution quickly converges to 
the \v{S}eba distribution \cite{seba1991}. 
This distribution does show a gap at small level spacings but
fails to reproduce the Gaussian tail predicted by the 
GOE.
At small $a_{s}$ the level spacing distribution tends to
the Poisson one (see Fig.\ \ref{Fig_Volodya_level_statistics}). 
For the energy range $E \gtrsim 100 \hbar \omega_{\perp}$
and the aspect ratio $\left(L/a_{\perp }\right)^{2} = 2 \phi_{gr} \pi^7$
the \v{S}eba distribution is approached at $a_{s} \gtrsim 10^{-1}a_{\perp }$.

The level spacing statistics for the two-well system is consistent with the GOE prediction (Fig.\ \ref{Fig_Cavan_level_spacings_histogram}), 
even though the statistics is very poor. 

\subsection{Wavefunction statistics}
Another important measure of quantum chaos is the statistics of the wavefunction. 
Divide
the coordinate space of a quantum-chaotic system into regions larger that the 
de-Broglie wavelength but small enough so that the spatial density does not change 
substantially across the region.   
According to the Berry conjecture \cite{berry1977,heller1984},
the coordinate wavefunction should behave, within each region, as a sample from 
an ensemble of large superpositions of plane waves with random coefficients. 

For the waveguide system, we verified that in the globally unitary regime, $a_{s} \gg a_{\perp }$, the statistics of the point-by-point variations 
of the eigenstate wavefunction in the spatial representation is indeed close to the Gaussian one (Fig.\ \ref{Fig_Volodya_psi_berry_long}),
according to Berry's prediction and a particular observation 
in the case of the \v{S}eba billiard \cite{seba1990}.

The two-well system also behaves according to the Berry conjecture.
Fig.\ \ref{Fig_Cavan_psi_gaussian}a presents the distribution 
of the real part of the wavefunction (\ref{sphere}) taken at random points on the bi-sphere and averaged over 
the eigenstates between $\alpha = 150$ and $\alpha = 250$. For $J=2$ the distribution is indistinct from
the Gaussian. 
Fig.\ \ref{Fig_Cavan_psi_gaussian}b represents the dependence of the entropy of the distribution on the 
hopping constant. The entropy converges to the Berry conjecture prediction in the region
between $J/U \approx 1$ and $J/U \approx 5$.

\begin{figure}
\includegraphics[width=4.5in]{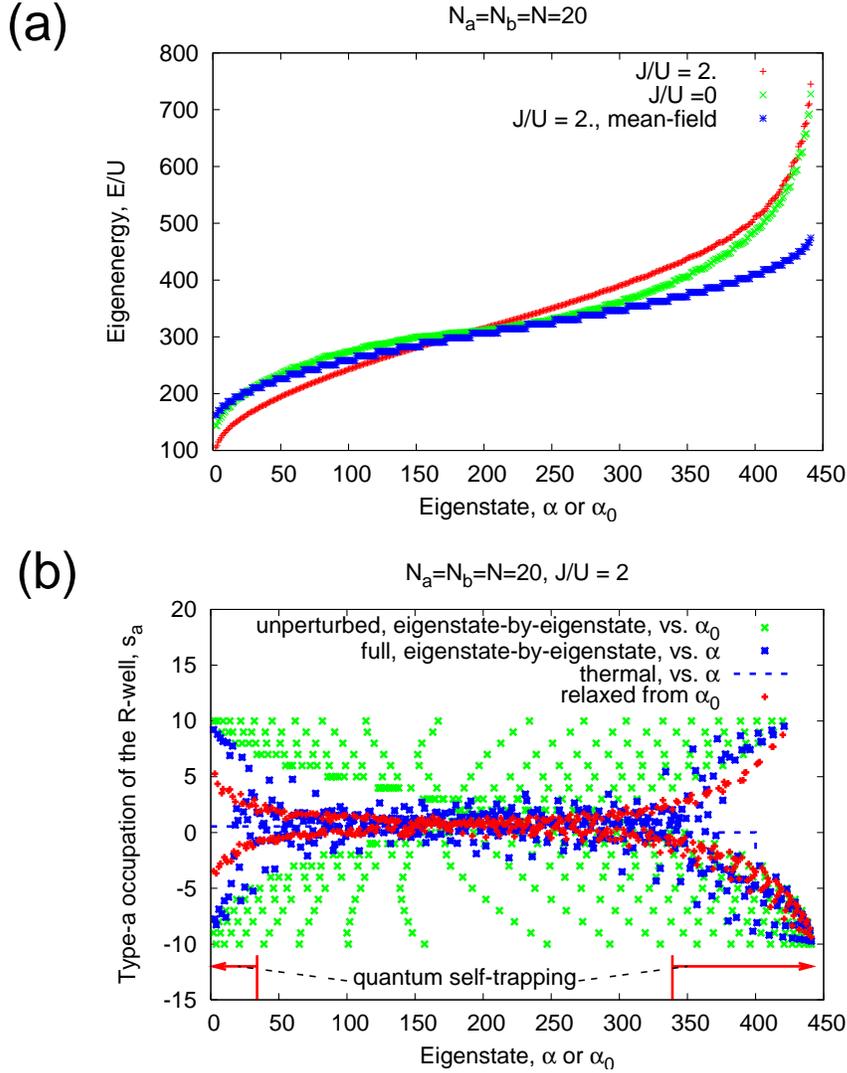}
\caption
{
Forty bosons in two wells.
(a) Energy spectrum of the Hamiltonian (\ref{two_well_hamiltonian}) in the absence of hopping ($J/U=0$)
and for the hopping $J/U=2$. The spectrum of the mean-field Hamiltonian 
$\hat{H}_{\mbox{\scriptsize mf}} \equiv \hat{V} + \langle \hat{H}_{0} \rangle$ (see (\ref{two_well_Hmf}), with the hopping as the 
primary Hamiltonian, is shown as well. The point $\alpha \approx 190$ where all three eigenenergy curves 
coincide corresponds to the maximal depolarization where the motion is ``maximally chaotic''.
Here, the density is distributed uniformly over the bi-sphere with no preferred orientation.
The labels $\alpha$ and $\alpha_{0}$ correspond to the eigenstates of the full Hamiltonian 
(\ref{two_well_hamiltonian}) and the pure-interaction Hamiltonian ($J/U=0$ limit) respectively.
(b) Fluctuations of the observable $\hat{s}_{a} \equiv \hat{c}^{\dagger}_{aR} \hat{c}^{}_{aR} - N_{a}/2$.
Green crosses correspond to the microcanonical fluctuations in the absence of hopping. Blue crosses correspond to the 
eigenstate-by-eigenstate fluctuations of the quantum expectation values, suppressed 
according to the eigenstate thermalization hypothesis. Red crosses give the infinite time average 
of the quantum expectation for a propagation from an initial state corresponding to one of the 
eigenstates of the Hamiltonian $\hat{H}_{0}$. The fluctuations of the latter are further 
suppressed due to an additional averaging of the initial state over the eigenstates. The microcanonical 
average is shown as the blue dashed line.
}
\label{Fig_cavan_spectra_and_eth}
\end{figure}

Fig.\ \ref{Fig_Cavan_psi_sphere}  
shows the wavefunction of the eigenstate $\alpha = 190$ at hopping of $J=2$. This state corresponds to the crossing of all three spectra
of the Fig.\ \ref{Fig_cavan_spectra_and_eth}a where the expectation values of the projection 
of the angular momentum $\hat{\vec{\cal L}}_{a}$ (along with $\hat{\vec{\cal L}}_{b}$) on both $z$ and $x$ axes is small. 
Accordingly, the nodal structure of the wavefunction appears completely random, with no preferred direction.
For comparison, Fig.\ \ref{Fig_Cavan_psi_sphere_noninteracting} shows the same state but for weak hopping ($J=.2$)
where the motion is expected to be regular.

\begin{figure}
\includegraphics[width=6.5in]{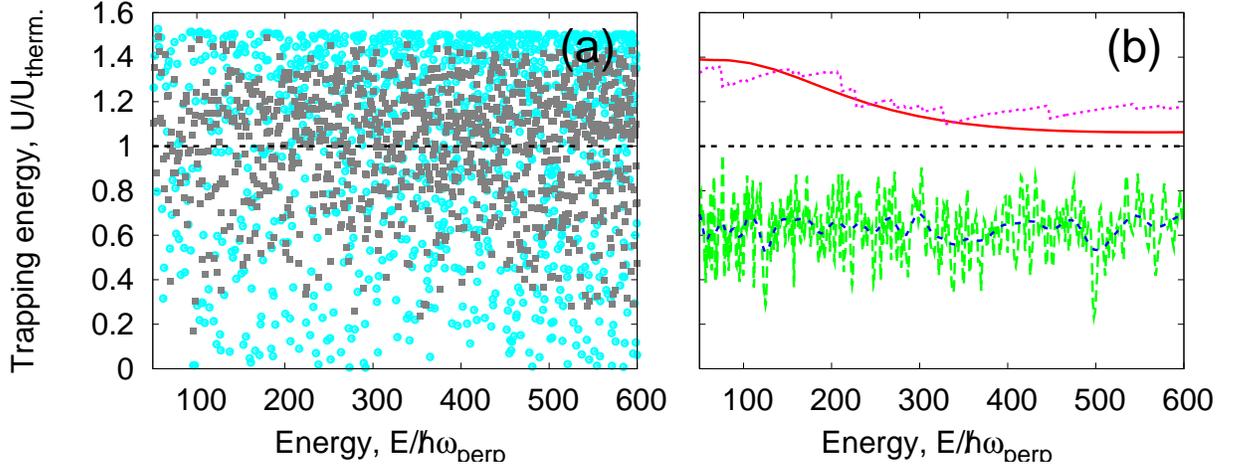}
\caption
{
Two bosons in a multimode guide.
(a) 
Quantum expectation value of
the trapping energy $U$ for every hundreds eigenstate of the
unperturbed Hamiltonian (\ref{H0_2b}) (blue circles) and the full Hamiltonian (\ref{H_2b}) (grey squares).
(b) 
The time average of the transverse trapping energy
$U_{\mbox{\scriptsize relax.}} \equiv \lim_{t_{max}\to\infty} (1/t_{max}) 
\int_{0}^{t_{max}}\! \langle \psi(t) | \hat{U} | \psi(t) \rangle$
after relaxation from an initial state as a function
of the energy of the initial state $E$ [see Hamiltonian (\ref{H_2b})]. Four families 
of the initial states are considered. Long-dashed (green) line:
state (\protect\ref{pdn}) with $n_{0}=0$, $\delta=0.99$,
and the energy being controlled via scanning $l_{0}$;
short-dashed (blue) line: the same as the previous one, except 
for $\delta=.1$; dotted (purple) line: the state (\protect\ref{pdn})
with $\delta=0.1$ and $l_{0}=0$ and the energy control via $n_{0}$; 
solid (red) line: state (\protect\ref{pdkk}) with
$\kappa_{2}=2\kappa _{1}$, $\delta=0.99$, $l_{0}=0$ and 
the energy control via $\kappa_1$. 
All the values of the trapping energy are shown with respect to its 
microcanonical expectation value $U_{therm.} \approx E/3$. 
The rest of the parameters is the same as  at Fig.\ \ref{Fig_Volodya_level_statistics}.
}
\label{Fig_csie}
\end{figure}

\begin{figure}
\includegraphics*[width=3.5in]{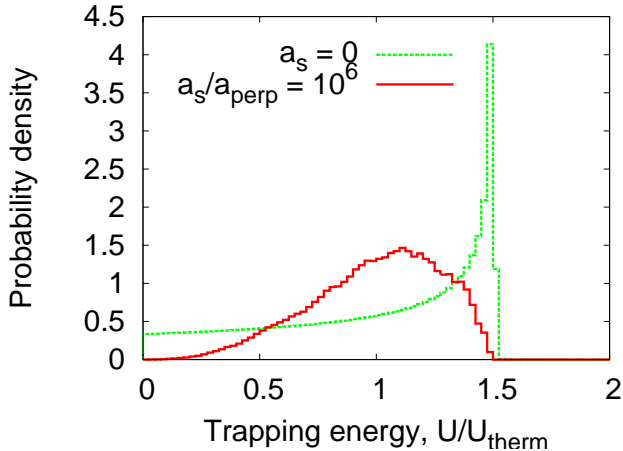}

\caption
{
Two bosons in a multimode guide. Probability distribution of the quantum expectation values of the trapping energy $U$
for individual eigenstates of the full Hamiltonian (\ref{H_2b}) in the globally unitary 
regime  (solid line) and those of the
unperturbed Hamiltonian (\ref{H0_2b}) (dashed line).
The rest of the parameters is the same as  at Fig.\ \ref{Fig_Volodya_level_statistics}.
}
\label{Fig_csidist}

\end{figure}

%




\section{Eigenstate thermalization phenomenon and the ability to thermalize in a relaxation from an initial state}
According to the Eigenstate Thermalization Hypothesis \cite{shnirelman1974,feingold1986,deutsch1991,srednicki1994},
the ability of an isolated quantum system to thermalize follows from suppression of the 
eigenstate-by-eigenstate fluctuations of the quantum expectation values of relevant observables 
$\langle \alpha | \hat{A} | \alpha \rangle$ over the eigenstates $| \alpha \rangle$ of the system.
Indeed, if one assumes that an observable $\hat{A}$ reaches its thermal value in a time evolution from 
any initial state with an energy close to a given energy $E$, {\it including} the exact eigenstates of the Hamiltonian, then  
it immediately follows that the matrix elements $\langle \alpha_{1} | \hat{A} | \alpha_{1} \rangle$ 
and $\langle \alpha_{2} | \hat{A} | \alpha_{2} \rangle$ must be close to each other if the corresponding 
energies $E_{1}$ and $E_{2}$ are close.

Consider first the system of two particles in a waveguide.
As the principal observable of interest, we choose the transverse trapping energy $U$
(see (\ref{H0_2b})). In our analysis, the potential energy $U$ plays a role of a generic observable 
in an integrable system: it acts on only one of the degrees of freedom and it has both diagonal and off-diagonal 
non-zero matrix elements between the eigenstates of the integrable system.

Scattered grey squares at the Fig.\ \ref{Fig_csie}a show the 
quantum expectation values of the transverse trapping energy, 
$\langle \alpha | \hat{U} | \alpha \rangle$, in every hundreds eigenstate 
(\ref{psieps}) of our 
system (\ref{H_2b}), relative to
its microcanonical expectation value $U_{therm.} \approx E/3$ (in accordance with the energy equipartition 
between the two harmonic and one free-space degrees of freedom).  

Even for the energies much larger than any conceivable energy scale of the system,
the variation of $\langle \alpha | \hat{U} | \alpha \rangle$ 
is only marginally lower than the microcanonical fluctuations in the unperturbed system (\ref{H0_2b}), represented by
the blue circles. 
Figure
\ref{Fig_csidist} demonstrates 
that even though the destruction of the integrals of motion 
in the interacting system does narrow the distribution of the diagonal matrix elements
$\langle \alpha | \hat{U} | \alpha \rangle$ and it does shift the peak towards
the microcanonical prediction $U_{therm.}$, the distribution remains relatively broad.
We trace this effect to the predicted unattainability of the complete quantum chaos in
the singular billiards \cite{seba1991,bogomolny2001,bogomolny2002}.

Here and below,
we assume the globally unitary value of the 
scattering length, 
$a_{s} = 10^{6} a_{\perp}$.

On the contrary, the behavior of the two-well system is fully consistent with the Eigenstate Thermalization Hypothesis
\cite{shnirelman1974,feingold1986,deutsch1991,srednicki1994} (see Fig.\ \ref{Fig_cavan_spectra_and_eth}b). 
Here, the observable $\hat{s}_{a}$ 
(the deviation number of the type-$a$ particles in the right well from $N/2$) is the primary observable of interest. 
In the window $150 \lesssim \alpha \lesssim 250$, all quantum expectation values of $\hat{s}_{a}$ in the eigenstates of the 
full Hamiltonian (\ref{two_well_hamiltonian}) are substantially closer to 
the thermal prediction than their unperturbed counterparts given by the eigenstates $|\alpha_{0}\rangle = |s_{a0},\,s_{b0}\rangle$ of the Hamiltonian (\ref{two_well_H0}).  




Now we are going to address directly the ability of our system to thermalize 
from an initial state $|\psi(t\!=\!0)\rangle$. In this case,
the infinite time average of the quantum expectation value of an observable 
$\hat{A}$ will be given by 
\begin{eqnarray*}
A_{\mbox{\scriptsize relax.}} &\equiv& \lim_{t_{max}\to\infty} (1/t_{max}) 
\int_{0}^{t_{max}}\! \langle \psi(t) | \hat{A} | \psi(t) \rangle 
\\
&=&
\sum_{\alpha} |\langle \alpha | \psi(t\!=\!0)\rangle|^2 \langle \alpha | \hat{A} | \alpha \rangle
\quad, 
\end{eqnarray*} 
where $|\alpha\rangle$ are the eigenstates of the system (see \cite{srednicki1994}).

First, let us attract our attention to the waveguide system.

Consider the following initial state:
\begin{eqnarray}
%
\langle z |\psi(t\!=\!0)\rangle 
=C_{\mbox{\scriptsize ax}}
\cos{\pi \zeta \over \delta}\theta \left(
 {\delta\over 2}-|\zeta |\right) \cos\left(2\pi l_{0}\zeta \right)
|n_{0}\rangle  
\label{pdn}
\\
C_{\mbox{\scriptsize ax}}=L^{-1/2}\left\lbrack {\delta\over 4}+{1\over 8\pi l_{0}} {\pi ^{2}\sin\left
( 2\pi l_{0}\delta\right) \over \pi ^{2}-\left( 2\pi l_{0}\delta\right) ^{2}}\right\rbrack ^{-1/2} \nonumber
\,.
\end{eqnarray}
Longitudinally, the state is represented by the ground state 
of a length 
$L\delta$ 
hard-wall box split initially by an ideal 
beamsplitter with momenta $\pm 2\pi l_{0}/L$; the box is
centered at the 
maximal interatomic distance.
The state is distributed among approximately  $\pi/\delta$ axial modes localized 
about $l=\pm l_{0}$.
The initial 
transverse state is limited to a single mode $n_{0}$. 
Note that for $n_{0} = 0$ and $\delta \approx 1$ the state 
is conceptually similar to the initial state used in the equilibration experiments \cite{kinoshita2006}.  

The transverse trapping energy $U$ 
(see (\ref{H0_2b}))
remains very far from the equilibrium predictions after the relaxation.
Fig.\ \ref{Fig_csie}b shows the equilibrium value of the transverse trapping energy $U$
for three sequences of the initial states of the type (\ref{pdn}). For the first two,
the energy is controlled via the kinetic energy given by the beamsplitter, while the 
transverse state is fixed to the ground state. Both 
show an approximately -40\% deviation of the relaxed value of $U$ from 
the thermal prediction. Note that the initial states of the second sequence
have a much greater spread over the longitudinal modes than the first one; consistently, the 
energy-to-energy variation  
of the relaxed values of $U$ is less than for the first sequence. In the third sequence,
there is no beamsplitting, and the energy is controlled via the initial transverse energy.
In this case, the deviation from the thermal prediction ranges between +50\% 
and +5\% but 
never reaches zero. 

Another type of the initial state
\begin{eqnarray}
\langle \rho,\,z | \psi(t\!=\!0) \rangle
&=&C_{\mbox{\scriptsize ax}}
\cos{\pi \zeta \over
 \delta}\theta \left( {\delta\over 2}-|\zeta |\right) \cos\left(2\pi l_{0}\zeta
 \right) \nonumber 
\\
&&
\times{1\over a{ } _{\perp }}\left\lbrack {2\kappa _{1}\kappa
 _{2}\left( \kappa _{1}+\kappa _{2}\right) \over \pi \left( \kappa
 _{1}-\kappa _{2}\right) { } ^{2}}\right\rbrack ^{1/2}
 \left(e^{-\kappa_1\xi }-e^{-\kappa_2\xi }\right)  
\label{pdkk}
\end{eqnarray}
allows for an additional spread over the transverse modes. 
The transverse wavefunction vanishes at the
waveguide axis. The mode occupation has a minimum at $n=0$, and the
occupation of the ground transverse mode tends to zero when $\kappa_{1} < \kappa _{2}\ll  1$.
The results for a sequence similar to the third sequence described above are 
shown at Fig.\ \ref{Fig_csie}b as well. In spite of the significant 
differences in the
initial distribution 
over both the transverse and the longitudinal modes, the deviations from the equilibrium 
are close to the ones for the third sequence, even though the energy-to-energy variations 
for the former are less than for the latter.

For our second system, we use the eigenstates of the unperturbed Hamiltonian (\ref{two_well_H0}) as the initial 
states of the relaxation process. The results are shown at Fig.\ \ref{Fig_cavan_spectra_and_eth}b. Here, 
each red cross shows the infinite time average of the quantum expectation value of the observable 
$\hat{s}_{a}$ (the deviation of the type-$a$ occupation of the right well from $N_{a}/2$)
for a relaxation process originating from the state $|\alpha_{0}\rangle = |s_{a0},\,s_{b0}\rangle$. 
In the window $150 \lesssim \alpha \lesssim 250$ the observable thermalizes fully. Recall, that 
the eigenstate themalization occurs in the same window.

It is instructive to analyze our results from the point of view of the macroscopic quantum self-trapping, 
predicted in Refs. \cite{milburn1997,smerzi1997,raghavan1999} and experimentally observed 
in Ref. \cite{albiez2005}, for the single-specie two-well bosonic system. From the classical field 
theory point of view it is a one-dimensional Hamiltonian system similar to a pendulum. Here, the population 
imbalance between the two wells plays a role of momentum, while the relative phase between the 
wells plays a role of a coordinate. Similarly to the case of the conventional pendulum,   
the initial states of sufficiently high momentum preserve the sign of the latter; the time evolution 
that starts from the states with 
even higher momentum shows the momentum that almost does not change in time. The states with the opposite 
sign of the momentum become mutually inaccessible. In terms of the original two-well
system, the initial states with the population imbalance higher than a certain critical value preserve the imbalance 
over time.

The energy landscape for out two-specie two-well system is more complicated. Both the cosine-like dependence of the hopping 
energy on the relative phases and the hyperbolic shape of the interaction energy surface as a function of the two
population imbalances contribute to the potential multi-connectedness of the phase space. At the same time, the initial 
states used to produce the Fig.\ \ref{Fig_cavan_spectra_and_eth}b  are the exact eigenstates of both population 
imbalances; they leave the relative phases completely 
uncertain.

We have identified a set of {\it absolutely localized} pairs of population imbalances, i.e.\ the pairs for which 
no set of imbalances of opposite signature is classically accessible for any pair of initial phases. In particular,
the analysis shows that all the unperturbed eigenstates with indices $\alpha_{0} < 39$ or $\alpha_{0} > 339$ constitute 
the absolutely localized initial states. This result is 
consistent with the fully quantum calculation that shows a 
strong memory of the 
initial conditions (see Fig.\ \ref{Fig_cavan_spectra_and_eth}b).    




\section{Summary of results}
In this article, we analyze the behavior of two simple atomic 
systems within two complementary frameworks: quantum chaos and nonequilibrium
dynamics. The first system is represented by two short-range-interacting bosonic 
atoms in a circular, transversely harmonic multimode waveguide. The second is the 
interacting two-component Bose-Bose mixture in two coupled potential wells. For both systems,
we study the level spacing statistics, the statistics of the values of the wavefunction in 
coordinate representation, degree of the eigenstate thermalization, and the 
thermalizability in a relaxation from an exited initial state.

For the waveguide system, the wavefunction statistics is fully 
consistent with the Berry conjecture \cite{berry1977} that
predicts a Gaussian distribution for the quantum-chaotic systems. 
However, the rest of the tests show an incomplete chaotization.
We trace this effect to the previously predicted incomplete 
quantum chaos in the singular billiards \cite{seba1991,albeverio1991,bogomolny2001,berkolaiko2003}.

For the double-well system, we identify a quantum-chaotic region of the spectrum that spans, for our set 
of parameters, about a quarter of the full spectrum. In this part of the spectrum,
the system exhibits both strong eigenstate thermalization and the ability to fully thermalize from 
an initial state.




\ack

We are grateful to Felix Werner for the enlightening discussions on the subject. 
This work was supported by a grant from the Office of Naval Research ({\it N00014-09-1-0502}).



%

\section*{References}

\bibliographystyle{unsrt}
\bibliography{therm_memory}


\end{document}